\documentstyle[twoside,fleqn,npb,epsfig,axodraw]{article}
\def\nl{\nonumber\\}

\def\nln{\nonumber\\*[-1ex]\phantom{\fbox{\rule{0em}{2ex}}}}

\newcommand{\gsim}
{\mathrel{\raisebox{-.3em}{$\stackrel{\displaystyle >}{\sim}$}}}
\def\asymp#1%
{\mathrel{\raisebox{-.4em}{$\widetilde{\scriptstyle #1}$}}}

\def\Nequal#1%
{\mathrel{\raisebox{-.5em}{$\stackrel{=}{\scriptstyle\rm#1}$}}}
\newcommand{\dsl}[1]{\not \hspace{-0.7mm}#1}
\def\dsl{\mathpalette\make@slash}
\def\make@slash#1#2{\setbox\z@\hbox{$#1#2$}%
  \hbox to 0pt{\hss$#1/$\hss\kern-\wd0}\box0}

\def\beq{\begin{equation}}
\def\eeq{\end{equation}}
\def\beqar{\begin{eqnarray}}
\def\eeqar{\end{eqnarray}}
\def\barr#1{\begin{array}{#1}}
\def\earr{\end{array}}
\def\bfi{\begin{figure}}
\def\efi{\end{figure}}
\def\btab{\begin{table}}
\def\etab{\end{table}}
\def\bce{\begin{center}}
\def\ece{\end{center}}
\def\nn{\nonumber}

\def\text{\textstyle}

\def\de{\delta}

\def\la{\lambda}
\def\si{\sigma}

\def\refeq#1{\mbox{(\ref{#1})}}
\def\reffi#1{\mbox{Fig.~\ref{#1}}}

\def\citere#1{\mbox{Ref.~\cite{#1}}}
\def\citeres#1{\mbox{Refs.~\cite{#1}}}

\def\solid{\raise.9mm\hbox{\protect\rule{1.1cm}{.2mm}}}
\def\dash{\raise.9mm\hbox{\protect\rule{2mm}{.2mm}}\hspace*{1mm}}

\newcommand{\TeV}{\unskip\,\mathrm{TeV}}

\def\mathswitchr#1{\relax\ifmmode{\mathrm{#1}}\else$\mathrm{#1}$\fi}

\newcommand{\PW}{\mathswitchr W}
\newcommand{\PZ}{\mathswitchr Z}

\newcommand{\PH}{\mathswitchr H}

\newcommand{\Pf}{\mathswitchr f}

\newcommand{\Pt}{\mathswitchr t}

\newcommand{\Pep}{\mathswitchr {e^+}}
\newcommand{\Pem}{\mathswitchr {e^-}}

\newcommand{\PWpm}{\mathswitchr {W^\pm}}

\def\mathswitch#1{\relax\ifmmode#1\else$#1$\fi}

\newcommand{\MW}{\mathswitch {M_\PW}}

\newcommand{\MZ}{\mathswitch {M_\PZ}}
\newcommand{\MH}{\mathswitch {M_\PH}}

\newcommand{\Mt}{\mathswitch {m_\Pt}}

\newcommand{\scrs}{\scriptscriptstyle}
\newcommand{\sw}{\mathswitch {s_{\scrs\PW}}}
\newcommand{\cw}{\mathswitch {c_{\scrs\PW}}}

\newcommand{\cew}{C^{\ew}}

\def\ie{i.e.\ }

\newcommand{\etal}{{\it et al.}}

\renewcommand{\O}{{\cal O}}

\newcommand{\EW}{{\mathrm{EW}}}
\newcommand{\sem}{{\mathrm{sem}}}
\newcommand{\sew}{{\mathrm{sew}}}

\newcommand{\SUtwo}{\mathrm{SU}(2)}
\newcommand{\Uone}{\mathrm{U}(1)}

\newcommand{\LL}{\mathrm{LL}}
\newcommand{\NLL}{\mathrm{NLL}}

\newcommand{\rd}{{\mathrm{d}}}

\newcommand{\ew}{\mathrm{ew}}

\newcommand{\M}{{\cal {M}}}

\newcommand{\diagtwoL}[2]{A_{#1}^{#2}}
\newcommand{\diagtwoC}[2]{B_{#1}^{#2}}
\newcommand{\diagtwoY}[2]{C_{#1}^{#2}}
\newcommand{\diagthreeL}[2]{D_{#1}^{#2}}
\newcommand{\diagthreeY}[2]{E_{#1}^{#2}}
\newcommand{\diagfourL}[2]{F_{#1}^{#2}}

\newcommand{\leg}[1]{{#1}}

\newcommand{\blob}{
\Vertex(-15.9138,3.75675){0.8}
\Vertex(-16.3512,-0.00758122){0.8}
\Vertex(-15.9138,-3.75675){0.8}
\Line(0.,0.)(-19.5,-9.75)
\Line(0.,0.)(-19.5,9.75)
\GCirc(0.,0.){10.9008}{0.5}
}

\newcommand{\smalldiagramtwoL}{
\begin{picture}(69.75,60.45)(-16.275,-30.225)
\Line(0.,0.)(46.5,23.25)
\Line(0.,0.)(46.5,-23.25)
\Photon(37.2,18.6)(37.2,-18.6){3}{3.5}
\Photon(18.6,9.3)(18.6,-9.3){3}{2}
\Vertex(37.2,18.6){2}
\Vertex(37.2,-18.6){2}
\Vertex(18.6,9.3){2}
\Vertex(18.6,-9.3){2}
\Text(48.825,24.4125)[l]{$\leg{j}$}
\Text(48.825,-24.4125)[l]{$\leg{k}$}
\Text(41.5909,0.)[l]{$a$}
\Text(20.7954,0.)[l]{$b$}
\Text(27.5691,14.593)[br]{}
\Text(27.9,-13.95)[tr]{}
\Text(13.6117,7.61406)[br]{}
\Text(13.95,-6.975)[tr]{}
\blob
\end{picture}
}

\newcommand{\smalldiagramtwoC}{
\begin{picture}(69.75,60.45)(-16.275,-30.225)
\Line(0.,0.)(46.5,23.25)
\Line(0.,0.)(46.5,-23.25)
\Photon(37.2,18.6)(18.6,-9.3){3}{3}
\Photon(37.2,-18.6)(18.6,9.3){3}{3}
\Vertex(37.2,18.6){2}
\Vertex(37.2,-18.6){2}
\Vertex(18.6,9.3){2}
\Vertex(18.6,-9.3){2}
\Text(48.825,24.4125)[l]{$\leg{j}$}
\Text(48.825,-24.4125)[l]{$\leg{k}$}
\Text(33.2131,6.23127)[l]{$a$}
\Text(33.2131,-6.23127)[l]{$b$}
\Text(27.5691,14.593)[br]{}
\Text(27.9,-13.95)[tr]{}
\Text(13.6117,7.61406)[br]{}
\Text(13.95,-6.975)[tr]{}
\blob
\end{picture}
}

\newcommand{\smalldiagramtwoY}{
\begin{picture}(69.75,60.45)(-16.275,-30.225)
\Line(0.,0.)(46.5,23.25)
\Line(0.,0.)(46.5,-23.25)
\Photon(37.2,18.6)(31.1931,0.){3}{2}
\Photon(18.6,9.3)(31.1931,0.){3}{2}
\Photon(31.1931,0.)(31.1931,-15.5966){3}{2}
\Vertex(37.2,18.6){2}
\Vertex(18.6,9.3){2}
\Vertex(31.1931,0.){2}
\Vertex(31.1931,-15.5966){2}
\Text(48.825,24.4125)[l]{$\leg{j}$}
\Text(48.825,-24.4125)[l]{$\leg{k}$}
\Text(37.9484,8.9584)[l]{$a$}
\Text(35.4185,-8.36117)[l]{$c$}
\Text(23.0875,3.7798)[tr]{$b$}
\Text(27.5691,14.593)[br]{}
\Text(20.925,-10.4625)[tr]{}
\Text(13.6117,7.61406)[br]{}
\blob
\end{picture}
}

\newcommand{\smalldiagramthreeLinv}{
\begin{picture}(69.75,60.45)(-16.275,-30.225)
\Line(0.,0.)(46.5,23.25)
\Line(0.,0.)(46.5,-23.25)
\Line(0.,0.)(51.9886,0.)
\Photon(26.505,13.2525)(26.505,0.){3}{2}
\Photon(39.525,-19.7625)(39.525,0.){-3}{2.5}
\Vertex(26.505,13.2525){2}
\Vertex(39.525,-19.7625){2}
\Vertex(26.505,0.){2}
\Vertex(39.525,0.){2}
\Text(48.825,24.4125)[l]{$\leg{k}$}
\Text(54.588,0.)[l]{$\leg{j}$}
\Text(48.825,-24.4125)[l]{$\leg{l}$}
\Text(43.0082,-10.1528)[l]{$a$}
\Text(30.3587,7.16672)[l]{$b$}
\Text(33.2727,0.)[t]{}
\Text(22.9743,12.1608)[br]{}
\Text(16.275,-8.1375)[tr]{}
\Text(18.196,0.)[t]{}
\blob
\end{picture}
}

\newcommand{\smalldiagramthreeY}{
\begin{picture}(69.75,60.45)(-16.275,-30.225)
\Line(0.,0.)(46.5,23.25)
\Line(0.,0.)(46.5,-23.25)
\Line(0.,0.)(51.7652,-4.81397)
\Photon(32.55,16.275)(36.0405,5.04561){3}{2}
\Photon(27.9,-13.95)(36.0405,5.04561){-3}{3}
\Photon(45.0357,-4.18815)(36.0405,5.04561){3}{2}
\Vertex(32.55,16.275){2}
\Vertex(45.0357,-4.18815){2}
\Vertex(27.9,-13.95){2}
\Vertex(36.0405,5.04561){2}
\Text(48.825,24.4125)[l]{$\leg{j}$}
\Text(54.3535,-5.05467)[l]{$\leg{k}$}
\Text(48.825,-24.4125)[l]{$\leg{l}$}
\Text(37.2341,11.5738)[l]{$c$}
\Text(44.1219,2.45738)[l]{$b$}
\Text(32.4934,-9.27993)[l]{$a$}
\Text(18.6,-9.3)[tr]{}
\Text(20.6769,10.9447)[br]{}
\Text(20.7954,0.)[b]{}
\blob
\end{picture}
}

\newcommand{\smalldiagramfourL}{
\begin{picture}(69.75,60.45)(-16.275,-30.225)
\Line(0.,0.)(46.5,23.25)
\Line(0.,0.)(46.5,-23.25)
\Line(0.,0.)(51.7652,4.81397)
\Line(0.,0.)(51.7652,-4.81397)
\Photon(37.2,18.6)(41.4122,3.85118){3}{2}
\Photon(37.2,-18.6)(41.4122,-3.85118){-3}{2}
\Vertex(37.2,18.6){2}
\Vertex(37.2,-18.6){2}
\Vertex(41.4122,3.85118){2}
\Vertex(41.4122,-3.85118){2}
\Text(48.825,24.4125)[l]{$\leg{j}$}
\Text(54.3535,5.05467)[l]{$\leg{k}$}
\Text(54.3535,-5.05467)[l]{$\leg{l}$}
\Text(48.825,-24.4125)[l]{$\leg{m}$}
\Text(42.4914,12.1353)[l]{$a$}
\Text(42.4914,-12.1353)[l]{$b$}
\blob
\end{picture}
}

\title{Two-loop electroweak corrections at high energies}

\author{A.~Denner%
\address{Paul Scherrer Institut (PSI), CH-5232 Villigen, Switzerland}%
, 
M.~Melles$^{\mathrm{a}}$
and
S.~Pozzorini%
\address{Institut f\"ur Theoretische Teilchenphysik,
Universit\"at Karlsruhe,
D-76128 Karlsruhe, Germany}%
}
\begin{document}

\thispagestyle{empty}
\def\thefootnote{\fnsymbol{footnote}}
\setcounter{footnote}{1}
\null
\hfill  PSI-PR-02-20
\\
\strut\hfill TTP02-36\\
\strut\hfill hep-ph/0211196
\vskip 0cm
\vfill
\begin{center}
{\Large \bf
Two-loop electroweak corrections at high energies%
\footnote{
Talk presented  by S.~P. 
at the International Symposium on Radiative Corrections RADCOR 2002, September 8--13, Kloster Banz, Germany.
}
\par} \vskip 2.5em
{\large
{\sc A.~Denner$^1$, M.~Melles$^1$ and S.~Pozzorini$^2$
}\\[1ex]
{\normalsize $^1$ \it Paul Scherrer Institut\\
CH-5232 Villigen PSI, Switzerland}\\[2ex]
{\normalsize $^2$ \it
Institut f\"ur Theoretische Teilchenphysik,
Universit\"at Karlsruhe\\
D-76128 Karlsruhe, Germany
}}

\par \vskip 1em
\end{center}
\par
\vskip .0cm 
\vfill {\bf Abstract:} \par 
We discuss  two-loop leading and angular-dependent next-to-leading
logarithmic  electroweak virtual corrections to arbitrary processes 
at energies above the electroweak scale.
The relevant Feynman diagrams involving soft-collinear gauge 
bosons $\gamma,\PZ,\PW^\pm$ have been evaluated in eikonal approximation.
We present results 
obtained from the analytic evaluation of massive loop integrals.
To isolate mass singularities we used  the Sudakov method and alternatively the  sector decomposition method in the Feynman-parameter representation.
\par
\vskip 1cm
\noindent
November 2002 
\par
\null
\setcounter{page}{0}
\clearpage
\def\thefootnote{\arabic{footnote}}
\setcounter{footnote}{0}

\begin{abstract}
We discuss  two-loop leading and angular-dependent next-to-leading
logarithmic  electroweak virtual corrections to arbitrary processes 
at energies above the electroweak scale.
The relevant Feynman diagrams involving soft-collinear gauge 
bosons $\gamma,\PZ,\PW^\pm$ have been evaluated in eikonal approximation.
We present results 
obtained from the analytic evaluation of massive loop integrals.
To isolate mass singularities we used  the Sudakov method and alternatively the  sector decomposition method in the Feynman-parameter representation.
\end{abstract}

\maketitle

\section{Introduction}
The main task of future colliders such as the LHC
or an $\Pep\Pem$ Linear Collider (LC)
is  the investigation of the origin of electroweak symmetry breaking and the
exploration of the limits of the Electroweak Standard Model.  In order
to disentangle effects of physics beyond the Standard Model, the
inclusion of QCD and electroweak radiative corrections into the
theoretical predictions is crucial.

In the energy range of future colliders, \ie at energies above the
electroweak scale, $\sqrt{s}\gg\MW$, the electroweak
corrections are enhanced by large logarithmic contributions \cite{Kuroda:1991wn} of the type
\beq\label{logform}
\alpha^N\log^{M}{\left(\frac{s}{\MW^2}\right)},\quad M>0.
\eeq
The leading logarithms (LL), also known as Sudakov logarithms
\cite{Sudakov:1954sw}, correspond to $M=2N$,
the next-to-leading logarithms (NLL) to $M=2N-1$, etc.
The logarithmic dependence on various kinematic invariants 
$r=s,t,u,\dots$  gives rise to
subleading logarithms that involve ratios of invariants,
and  which we denote as angular-dependent logarithms
\beq\label{anglogform}
\alpha^N\log^{M-L}{\left(\frac{s}{\MW^2}\right)}\log^{L}{\left(\frac{|r|}{s}\right)},\quad M-L>0.
\eeq
We will consider the kinematical region $|r|\gg \MW^2$, where all invariants are much larger than the electroweak scale.
The general form  of electroweak logarithmic corrections 
is complicated by the 
hierarchy of mass scales 
$
\Mt\sim \MH\sim \MZ\sim \MW 
\gg m_f\gg \la,
$
with heavy masses 
at the electroweak scale, light-fermion masses $m_f$, and the photon mass $\la$  as infrared regulator.
As a consequence all logarithms of the large ratios ${\MW}/ m_f$ and ${\MW}/\la$  have  to be taken into account.

All the above logarithmic terms 
constitute the singular part of the corrections in the massless limit. 
They result  either 
as mass singularities
from soft/collinear  emission of virtual or real particles off initial or final-state particles,
or as remnant of ultraviolet singularities after parameter renormalization. 

At the one-loop level, it has been proven that for processes that are not mass-suppressed at high energies 
the electroweak logarithms are universal, and 
general results have been given \cite{Denner:2001jv} and applied to gauge-boson pair production at the LHC \cite{Accomando:2001fn}.
These results are in agreement with various explicit diagrammatic calculations for many $2\to2$ scattering processes 
\cite{Beenakker:1993tt,Ciafaloni:1999xg,Beccaria:2000fk,Beccaria:2001yf,Layssac:2001ur}.
The approximate size of the one-loop electroweak LL and NLL for a  typical 
$2\to 2$  cross section is %
\beqar\label{oneloopestimate}
\frac{\de \si_{1,\LL}}{\si_0} &\simeq&
-\frac{\alpha}{\pi\sw^2}\log^2\left(\frac{s}{\MW^2}\right)\simeq -26\%
,\nl
\frac{\de \si_{1,\NLL}}{\si_0} &\simeq&
+\frac{3\alpha}{\pi\sw^2}\log\left(\frac{s}{\MW^2}\right)\simeq 16\%,
\eeqar
at $\sqrt{s}=1\TeV$, with $1-\sw^2=\cw^2=\MW^2/\MZ^2$.
The LL and NLL have similar size and opposite sign resulting in large cancellations.

Assuming that at high energies the symmetric phase of the electroweak
theory can be used, 
resummations of the electroweak logarithms have
been proposed based on techniques and results known from QCD. 
Fadin \etal\ \cite{Fadin:2000bq} have resummed the LL by
means of the infrared evolution equation.
K\"uhn \etal\ have applied results from QCD to resum the logarithmic
corrections to massless 4-fermion processes, $\Pep\Pem\to\Pf\bar\Pf$
up to  the NLL \cite{Kuhn:2000nn} and even to the NNLL \cite{Kuhn:2001hz}.
It was found that at $1\TeV$ there is no
clear hierarchy between LL, NLL and NNLL,
and that the angular-dependent logarithms are important.
Melles has proposed a resummation of the NLL for arbitrary processes
 \cite{Melles:2001gw}, which relies on the prescription of matching a 
symmetric $\SUtwo\times\Uone$ theory with QED at the electroweak scale.
Recently also an extension of this resummation to  the angular-dependent 
NLL has been proposed \cite{Melles:2001dh}. 

All these resummations amount to exponentiations
of the electroweak logarithms. The approximate size of the 
resulting two-loop LL and NLL for typical $2\to 2$ processes 
at $\sqrt{s}=1\TeV$ is
\beqar\label{twoloopestimate}
\frac{\de \si_{2,\LL}}{\si_0} &\simeq&
+\frac{\alpha^2}{2\pi^2\sw^4}\log^4\left(\frac{s}{\MW^2}\right)\simeq 3.5\%
,\nl
\frac{\de \si_{2,\NLL}}{\si_0} &\simeq&
-\frac{3\alpha^2}{\pi^2\sw^4}\log^3\left(\frac{s}{\MW^2}\right)\simeq -4.2\%
,
\eeqar
and it is clear that in view of the precision
objectives of a LC below the per-cent level these two-loop
logarithms must be under control.

All the above resummation prescriptions result from 
matching a {\em symmetric} $\SUtwo\times\Uone$ theory and 
QED at the electroweak scale, assuming that other effects 
related to  spontaneous symmetry breaking may be neglected at high energies. 
This assumption needs to be checked by 
explicit diagrammatic two-loop calculations based on the  
electroweak Lagrangian, where  all nontrivial effects 
related to  {\em spontaneous symmetry breaking} are taken into account, 
in particular 
(i) the large gap $\MW\sim\MZ\gg\la$ in the gauge sector,
(ii) the mixing between the gauge-group eigenstates $B,W^3$ resulting into the mass eigenstates $\gamma,\PZ$,
and 
(iii) the presence of longitudinal gauge bosons as physical asymptotic states.

The resummation of the two-loop LL has been checked for the
massless fermionic form factor in \citeres{Melles:2000ed,Hori:2000tm} and for
arbitrary processes in the massive Coulomb gauge in \citere{Beenakker:2000kb}.
The resummation of the NLL has so far not been
confirmed by explicit electroweak two-loop calculations.

A subset of the NLL is furnished by the
angular-dependent logarithms of type \refeq{anglogform} with $M=2N$, $L=1$.
These contributions are numerically important 
\cite{Beccaria:2001yf,Kuhn:2001hz}. At
one-loop order, in the t'~Hooft--Feynman gauge,  they result only from diagrams where a gauge boson is
exchanged between two external lines. Similarly, the angular-dependent
NLL  at two-loop order can be traced back to a
relatively small set of Feynman diagrams. This allows us to perform 
a diagrammatic calculation of the two-loop angular-dependent
NLL for arbitrary processes. 
The relevant massive two-loop
integrals have been evaluated in  eikonal approximation, and the 
logarithms have been obtained analytically using
two independent methods: the first one goes back to Sudakov
\cite{Sudakov:1954sw}, the other one uses sector decomposition
 of Feynman-parameter integrals \cite{Hepp:1966eg,Roth:1996pd,Binoth:2000ps}.
A detailed description of this calculation can be found in \citere{nextpaper}.
Here we summarize the main ingredients and results.


\begin{figure*}
\beqar\label{diagtwolegs}
\diagtwoL{jk}{ab}&=&
\vcenter{\hbox{\smalldiagramtwoL}}
\, ,\qquad
\diagtwoC{jk}{ab}=
\vcenter{\hbox{\smalldiagramtwoC}}
\, ,\qquad
\diagtwoY{jk}{abc}=
\vcenter{\hbox{\smalldiagramtwoY}}\, ,\nl
\diagthreeL{jkl}{ab}&=&
\vcenter{\hbox{\smalldiagramthreeLinv}}
\quad,\qquad
\diagthreeY{jkl}{abc}=
\vcenter{\hbox{\smalldiagramthreeY}}
\quad,\qquad
\diagfourL{jklm}{ab}=
\vcenter{\hbox{\smalldiagramfourL}}
\quad.\nn
\eeqar
\vspace{3mm}

{Figure 1. Two-loop  diagrams with soft-collinear gauge bosons $a,b,c=\gamma,\PZ,\PWpm$ exchanged between external legs $j,k,l,m=1,\dots,n$.}
\label{diagrams}
\end{figure*}

\section{Feynman diagrams}
In the following we consider electroweak processes%
\footnote{As a convention, all particles $\varphi_{i_k}$  and their momenta $p_k$ are assumed to be incoming. The corresponding $2\to n-2$ processes are easily obtained by crossing symmetry.}
$
\varphi_{i_1}(p_1)\ldots \varphi_{i_n}(p_n)\to 0
$,
involving $n$
arbitrary mass-eigenstate  particles.   
The kinematical invariants 
are denoted as $r_{kl}=(p_k+p_l)^2\simeq 2p_kp_l$, and the  matrix elements as
\beq\label{mel}
\M \equiv \M^{\varphi_{i_1}\ldots \varphi_{i_n}}(p_1,\dots,p_n).
\eeq
We restrict ourselves to matrix elements that are not mass-suppressed at high energies.  In this case global gauge invariance implies
\beq\label{chargecons}
\sum_{k=1}^n  \M\, I^a (k) = \O\left(\frac{M^2}{s}
\right)\, \M,\quad a=\gamma,\PZ,\PWpm,
\eeq
where the gauge couplings $I^a(k)$ act as (transposed) matrices on the 
external-legs 
$k$ of the matrix element%
\footnote{Details concerning our notation can be found in \citere{Denner:2001jv}.}.

In the t'~Hooft--Feynman  gauge, the leading mass singularities  originate 
from diagrams with soft-collinear  virtual gauge bosons  coupling to external particles.
The relevant two-loop diagrams are depicted in \reffi{diagrams}, where 
the  soft-collinear gauge bosons 
are  exchanged between two, 
three, 
or four 
of the $n$ on-shell external legs.

Each loop integral has to be evaluated for all different mass assignments 
that  occur in the electroweak model. For the internal lines we have the cases $a,b,c=\gamma,\PZ,\PWpm$ and 
for the external masses we assume $\MW\gsim m_{\mathrm{ext}}\gg \la$.

The Feynman diagrams are evaluated in  eikonal approximation, \ie by 
neglecting mass terms and  the momenta of the soft gauge bosons everywhere in the numerators, apart from the momenta 
in the couplings of three soft gauge bosons.
In the massless limit, longitudinal gauge bosons have to be  substituted by the corresponding would-be Goldstone Bosons using the 
Goldstone-Boson equivalence theorem.

\section{Loop integrals in logarithmic approximation}

In the  evaluation of the loop integrals we include the LL and the  angular-dependent NLL, and we use  $\Mt\simeq \MH \simeq \MZ \simeq \MW$, \ie we neglect logarithms of ratios of heavy masses.
Two analytical methods have been used:
the Sudakov method \cite{Sudakov:1954sw} 
 and sector decomposition \cite{Hepp:1966eg,Roth:1996pd,Binoth:2000ps}, which 
permits to factorize  overlapping ultraviolet or mass singularities in Feynman-parameter integrals.
Here we only sketch the main steps of the second method         applied to a generic two-loop massive integral.

{\bf Step 1:} The integral
is written in Feynman parametrization
and the denominator is split into polynomials  according to the hierarchy of scales $s\gg r \gg M^2\gg\dots\gg\la^2$ in the diagram\footnote{To extract the angular-dependent logarithms $\log{(s/r)}$ with  $r=t,u,\dots$, we compute the integrals 
in the euclidean region 
in various limits of the type $s\gg t=u$, $s=t\gg u$, etc., where we separate the energy scales in various ways.}
\beqar
I&=&
\int_{[0,1]^n} \rd \vec{x}\,
\frac{
f(\vec{x})}{\left[
D(\vec{x})
\right]^e}
,\\
D(\vec{x})&=&{s} P_{s}(\vec{x})+{r} P_{r}(\vec{x})+\ldots +{\la^2} P_{\la}(\vec{x}).\nn
\eeqar
These polynomials have various zeros 
of the form 
\beq\label{polyone}
P(\vec{x})=\sum_{i=1}^m x_i P_i(\vec{x})=0,
\eeq
at $x_1=\dots=x_m=0$, which give rise to  mass singularities.

{\bf Step 2:} In order to separate  overlapping singularities, we  decompose  the sector  $[0,1]^m$ into $m$ subsectors $\Omega_j$ with $x_j>x_{i\neq j}$, and in each subsector $\Omega_j$ we perform variable transformations $x_i\to x_j x'_i$, which remap $\Omega_j\to[0,1]^m$ and permit to factorize the variable  $x_j$ in
\beq\label{polytwo}
P(\vec{x})= \left[ P_j(\vec{x})+\sum_{i\neq j} x'_i P_i(\vec{x})\right]{x_j}.
\eeq

{\bf Step 3:} Recursive application of step 2 permits to factorize all zeros
at all scales, until  the 
denominator assumes the form 
\beqar
D(\vec{x})&=&
\left\{\left[{s} \hat{P}_s(\vec{x})\prod_k {x_k}
+{r} \hat{P}_r(\vec{x})\right]\prod_l {x_l}
\right.\nl&&\left.{}
+\ldots +{\la^2} \hat{P}_\la(\vec{x})\right\}\prod_m {x_m},
\eeqar
where all Feynman parameters that give rise to mass singularities  are factorized and the polynomials $\hat{P}$ are non-vanishing. This allows for 
a simple power counting of the logarithmically divergent integrations.

{\bf Step 4:} 
In leading-logarithmic approximation  the polynomials $\hat{P}$  can be treated as constants $\hat{P}(\vec{x})\simeq \hat{P}(\vec{0})$, and 
all logarithms of ratios of scales can be extracted by analytical integration of the singular parameters. 

\section{Results}
As a basis for the  presentation of our two-loop results, we recall the one-loop results 
for LL and angular-dependent NLL given in \citere{Denner:2001jv}.
\subsection{One-loop results}
At one-loop level we have
\beq
\M_1= \M_{0}(1+ \de_\EW ).
\eeq
The most symmetric form to write the   
electroweak logarithmic corrections consists in splitting them  into 
$\de_\EW=\de_\sew+\de_\sem$, with
\beqar\label{splitting}
\de_\sew
&=&
\left.\de_\EW\right|_{\la=\MW}
,\nl
\de_\sem
&=&
\left. \de_{\EW} -\de_{\EW}\right|_{\la=\MW}.
\eeqar
The symmetric electroweak part (sew) corresponds to the case when the photon mass $\la$ equals the electroweak scale and reads
\beqar\label{sewpart}
\lefteqn{
\de_\sew=
\frac{\alpha}{4\pi}\sum_{k=1}^n\left\{
-\frac{1}{2}\cew(k) \log^2\frac{s}{\MW^2}
\right.}
&&\\
&&\left.
\hspace{-7mm}{}
+\sum_{l\neq k} \sum_{a=\gamma,\PZ,\PWpm} 
I^a(k) I^{\bar{a}}(l)
\log\frac{|r_{kl}|}{s}
\log\frac{s}{\MW^2}
\right\},\nn
\eeqar
where 
$\cew=Y^2/(4\cw^2)+C^{\SUtwo}/\sw^2$ represents the electroweak Casimir operator.
The remaining part is a subtracted electromagnetic (sem) contribution 
originating from the fact that $\la\ll\MW$,
\beqar\label{sempart}
\lefteqn{
\de_\sem
=\frac{\alpha}{4\pi}\sum_{k=1}^n 
\left\{
-\frac{1}{2}Q^2(k)
\left[
2\log\frac{s}{m_k^2}\log{\frac{\MW^2}{\la^2}}
\right.\right.}
\quad&&\nl
&&\left.\left.
\hspace{-10mm}
-\log^2{\frac{\MW^2}{m_k^2}}
\right]
+\sum_{l\neq k}
Q(k) Q(l)
\log\frac{|r_{kl}|}{s}
\log\frac{\MW^2}{\la^2}
\right\}.\nln
\eeqar

\subsection{Two-loop results}
Detailed results for the individual two-loop diagrams in
\reffi{diagrams}
are presented in 
\citere{nextpaper}. 
These diagrams have to be combined as follows
\beqar\label{combination}
\lefteqn{
\de\M_2=\sum_{a,b}\left\{\sum_{j,k}
\left[\frac{1}{2}\left(\diagtwoL{jk}{ab}
+\diagtwoC{jk}{ab}\right)
+\sum_c \diagtwoY{jk}{abc}\right]
\right.
}&&
\nl&&\left.\hspace{-6mm}
{}+\sum_{j,k,l}
\left[\diagthreeL{jkl}{ab}
+\frac{1}{6}\sum_c\diagthreeY{jkl}{abc}
\right]
{}+\frac{1}{8}\sum_{j,k,l,m}
\diagfourL{jklm}{ab}\right\},\nn
\eeqar
taking into account all sums over virtual gauge bosons $a,b,c=\gamma,\PZ,\PWpm$ and external legs%
\footnote{In the sums only the contributions from different external legs  $j\neq k$, etc. have to be considered.}
$j,k,l,m=1,\dots,n$, with appropriate symmetry factors.
These sums can be simplified by means of \refeq{chargecons}, and it turns out
that the result corresponds to the exponentiation of the  one-loop corrections \refeq{sewpart},\refeq{sempart} in the form 
\beq\label{result}
\M_2=
\M_{0}
\exp\left(\de_\sew\right) 
\exp\left(\de_\sem\right), 
\eeq
where the symmetric electroweak and the subtracted electromagnetic parts exponentiate separately. 

\section{Discussion}
Our result confirms the exponentiation of the electroweak LL obtained with the infrared evolution equation \cite{Fadin:2000bq} and in the Coulomb gauge \cite{Beenakker:2000kb}.
The subset of diagrams (A-C) has been evaluated also in the special case of massless external particles and agreement has been found with the form-factor calculation of \citeres{Melles:2000ed,Hori:2000tm}. The exponentiation of the electroweak angular-dependent NLL 
agrees with the results of \citere{Kuhn:2001hz}
for massless fermionic processes, 
and \citere{Melles:2001dh} for  arbitrary processes. 
These results were obtained using  matching conditions at the electroweak scale. 

In the following we discuss the idea of matching by
applying it to our result. If we set $\la=\MW$ or $s=\MW^2$,  we obtain  
\beqar\label{matching}
&&\hspace{-6mm}
\la=\MW \Rightarrow \de_\sem =0,\,\,
\M_2=
\exp\left[\de_\sew\right] \M_{0},\nl
&&\hspace{-6mm}
s=\MW^2 \Rightarrow\de_\sew =0,\,\,
\M_2=\exp\left[\de_\sem\right] \M_{0},
\eeqar
\ie we observe  exponentiation 
within a  symmetric $\SUtwo\times\Uone$ theory ($\la=\MW$)
and QED ($s=\MW^2$). This provides a simple consistency check of our result. 
However, we stress that the matching conditions \refeq{matching}
are not sufficient in order to determine 
the  interference terms $\de_\sew \de_\sem$  between electroweak and electromagnetic contributions, which vanish at both matching points.
These two-loop terms are fixed by our 
full electroweak calculation ($s\gg\MW^2\gg\la^2$) and 
are crucial in order to predict the ordering of the two exponentials in \refeq{result},
which is non-trivial at the level of the 
angular-dependent NLL, since
\beq
[\de_\sew, \de_\sem]=\O\left[\log\frac{r_{kl}}{s}
\log^2\frac{s}{\MW^2}
\log\frac{\MW^2}{\la^2}
\right].
\eeq
The fact that $\de_\sem$ appears on the right-hand side in  \refeq{result}
means that the subtracted electromagnetic 
corrections 
are only sensitive to the electromagnetic charges of the external particles.

\end{document}